\documentclass[conference]{IEEEtran}
\IEEEoverridecommandlockouts
\usepackage{cite}
\usepackage{amsmath,amssymb,amsfonts}
\usepackage{algorithmic}
\usepackage{graphicx}
\usepackage{textcomp}
\usepackage{xcolor}
\usepackage[ruled]{algorithm2e}
\usepackage{bm}
\usepackage{amssymb}
\usepackage{multicol}
\usepackage{multirow} 
\def\BibTeX{{\rm B\kern-.05em{\sc i\kern-.025em b}\kern-.08em
    T\kern-.1667em\lower.7ex\hbox{E}\kern-.125emX}}
\begin{document}

\makeatletter
\newcommand{\rmnum}[1]{\romannumeral #1}
\newcommand{\Rmnum}[1]{\expandafter\@slowromancap\romannumeral #1@}
\makeatother

\title{Using EBGAN for Anomaly Intrusion Detection\\
\thanks{$^*$Corresponding Author}
}

\author{\IEEEauthorblockN{Yi Cui\textsuperscript{1}, Wenfeng Shen\textsuperscript{2,1,*}, Jian Zhang\textsuperscript{1}, Weijia Lu\textsuperscript{3}, Chuang Liu\textsuperscript{3}, Lin Sun\textsuperscript{3}, Si Chen\textsuperscript{4}}
\IEEEauthorblockA{\textsuperscript{1} School of Computer Engineering and Science, Shanghai University, Shanghai, China\\
\textsuperscript{2} Xi'an Innovation College, Yan’an University, Shaanxi, China\\
\textsuperscript{3} AI Lab, United Automotive Electronic Systems Co., Shanghai, China\\
\textsuperscript{4} School of Medicine, Shanghai University, Shanghai, China\\
Email: \{yicui, wfshen, zhangjian\}@shu.edu.cn,\{weijia.lu, chuang.liu, lin.sun\}@uaes.com,
\\Caroline-sisi-chen@hotmail.com\\
}
}

\maketitle

\begin{abstract}
As an active network security protection scheme, intrusion detection system (IDS) undertakes the important responsibility of detecting network attacks in the form of malicious network traffic. Intrusion detection technology is an important part of IDS. At present, many scholars have carried out extensive research on intrusion detection technology. However, developing an efficient intrusion detection method for massive network traffic data is still difficult. Since Generative Adversarial Networks (GANs) have powerful modeling capabilities for complex high-dimensional data, they provide new ideas for addressing this problem. In this paper, we put forward an EBGAN-based intrusion detection method, IDS-EBGAN, that classifies network records as normal traffic or malicious traffic. The generator in IDS-EBGAN is responsible for converting the original malicious network traffic in the training set into adversarial malicious examples. This is because we want to use adversarial learning to improve the ability of discriminator to detect malicious traffic. At the same time, the discriminator adopts Autoencoder model. During testing, IDS-EBGAN uses reconstruction error of discriminator to classify traffic records.
\end{abstract}


\section{Introduction}
In network security research, IDS has become an effective technique to protect networks or systems from attacks \cite{b1}. By monitoring network traffic, IDS is expected to issue alerts in time if malicious network traffic is identified \cite{b2}. The performance of IDS mainly depends on the quality of the intrusion detection technology, so intrusion detection technology is directly related to the detection rate of the whole system. In order to effectively deal with network attacks, many researchers have begun to study intrusion detection techniques.

In fact, intrusion detection is similar to the classification problem, which requires to determine whether the network traffic is normal or malicious. In classification problem, many machine learning algorithms have been applied to network anomaly detection and achieved good results. These algorithms perform classification tasks to identify malicious behaviors in network traffic, such as Support Vector Machine (SVM) \cite{b3} and K-Nearest Neighbors (KNN) \cite{b4}, etc. However, with the continuous upgrade of hacker attack methods and the massive amount of network traffic data faced, some traditional machine learning algorithms are not applicable for new intrusion detection scenarios.

In recent years, the advance of deep learning has also greatly promoted the development of intrusion detection. Different from traditional machine learning, deep learning learns the intrinsic laws of sample data, and the multi-layer nonlinear network structure constructed by deep learning can better adapt to the learning and prediction of complex high-dimensional data, which is more promising in solving intrusion detection problems \cite{b2}. Autoencoder (AE) \cite{b5}, Recurrent Neural Network (RNN) \cite{b6} and other deep learning algorithms are widely used for feature extraction tasks and sequence data processing tasks in intrusion detection.

However, network traffic data is not only complex and high-dimensional, but also has the problem of data class imbalance. Training an intrusion detection model with an unbalanced dataset where normal traffic data far outnumbers malicious traffic data may have a large impact on its accuracy. More importantly, the intrusion detection model exposes its weaknesses under the action of adversarial samples: for an already trained classification model, adding some imperceptible perturbations to malicious traffic samples in the training set will lead to incorrect classification results \cite{b7}. Based on the above problems, it is still a challenge to develop an efficient intrusion detection method.

GANs \cite{b8} are a powerful generative model that can model data distribution, which provides the possibility to solve the above difficulties. GANs consist of two modules which are used to reach an optimal solution through a minimax game \cite{b9}. As a model, GANs have been successfully applied in the fields of image \cite{b10}, natural language processing \cite{b11} and other fields. Current research focuses on using GAN for malicious traffic detection or generating adversarial malicious examples to improve the robustness of IDS.

Here we propose an EBGAN-based intrusion detection method IDS-EBGAN, which can detect malicious traffic accurately and efficiently. Inspired by IDSGAN \cite{b12}, the generator in IDS-EBGAN is used to modify the original malicious traffic to generate adversarial samples. It is worth noting that modifications to the original malicious traffic do not affect its attack capabilities. Based on the idea of sample reconstruction \cite{b13}, we try to use the EBGAN autoencoder model for intrusion detection tasks, where the discriminator reconstructs the input samples. The fidelity of the traffic sample reconstruction is used to measure whether the network traffic is anomalous or not. In addition, we combine state-of-the-art techniques \cite{b14} \cite{b15} to improve the encoder in the discriminator to stabilize the training of the model.

\section{Related Work}

\subsection{Traditional Methods}

In the field of academic research, there are numerous attempts to use traditional methods in intrusion detection tasks. Guan et al. \cite{b16} propose a clustering algorithm based on K-means and use it for intrusion detection. Terai et al. \cite{b17} introduce a SVM method for network attack detection in the industrial control domain. Their method realizes the discrimination between normal and malicious traffic by analyzing the features of network packets. In addition, decision tree algorithms \cite{b18} \cite{b19} are also widely used in intrusion detection.

\subsection{GAN-based Methods}
Lin et al. \cite{b12} propose the IDSGAN framework for adversarial attack sample generation against IDS. In later work, Usama et al. \cite{b20} add some constraints on the conditions for the generation of adversarial attack samples. In this paper, we incorporate the generation method of adversarial example into the intrusion detection method to improve the performance of the model.

Schlegl et al. \cite{b21} applied GAN to the task of eye image anomaly detection in 2017. During testing, AnoGAN needs to optimize the best latent vector $z$ corresponding to the test sample through the iterative process of the backpropagation algorithm. Since the backpropagation algorithm is time-consuming, it is not suitable for real-time network intrusion detection. With the proposal of BiGAN \cite{b22}, Zenati et al. propose the EGBAD model \cite{b23}, which solves the time-consuming problem of AnoGAN. Meanwhile, EGBAD has also achieved better results on the intrusion detection task. In 2018, the ALAD model was proposed \cite{b15}. It has two more discriminators and employs state-of-the-art techniques to further improve the performance of encoder and stabilize GAN training. In the same year, Akcay et al. \cite{b24} put forward the GANormaly model, which uses an encoder-decoder-encoder sub-network in the generator architecture. The above GAN-based anomaly detection models all reflect the idea of sample reconstruction. Inspired by existing research progress, we propose an EBGAN-based intrusion detection method.

\section{Background Knowledge}

\subsection{Adversarial Example}\label{AA}
First, some very small but purposeful perturbations are added to the samples in the dataset to form new input samples. Then feeding the perturbed samples into the model will result in the model outputting wrong answers with high confidence \cite{b25}. The sample with added perturbation is the adversarial sample. Adversarial training is a powerful method to enhance the robustness of neural networks.

\subsection{Energy-based Generative Adversarial Networks}
GANs were introduced by Goodfellow et al. in 2014 \cite{b8}. They consist of two neural networks, a generator $G$ and a discriminator $D$. $G$ converts random noise sampled from a Gaussian or uniform distribution into real samples, while $D$ evaluates the likelihood that the input samples come from $G$ or the real dataset. $G$ and $D$ play against each other. $G$ generates fake samples that are similar to real samples as much as possible, while $D$ maximizes the likelihood of assigning correct labels to the training samples and the samples generated by $G$. GANs are probabilistic-based models, which is a commonly accepted cognitive approach. Because the essence of $D$ is to calculate the conditional probability that the sample $x$ belongs to the category $y$, the essence of $G$ is to calculate the generation probability of the sample $x$ in the whole distribution.

EBGAN \cite{b26} is a GAN explained from an energy perspective, which treats the discriminator as an energy function. The discriminator $D$ is responsible for assigning low energy to high data density areas and high energy to those outside areas. The generator $G$ generates samples in the space where the discriminator assigns low energy. A typical example of the EBGAN framework is that the discriminator in EBGAN uses an AE architecture. And the reconstruction error of the AE in EBGAN is used as energy.

Suppose $p_{data}$ is the underlying probability density distribution of the generated the dataset. The generator $G$ first samples from a known distribution $p_z$, and then uses the extracted random vector $z$ to generate a sample $G(z)$. The discriminator $D$ takes generated samples or real samples as input, and evaluates the energy value $\mathbb{E}\in\mathbb{R}$ of each input sample. The architecture of the EBGAN autoencoder is presented in Fig. 1.

\begin{figure}[htbp]
\centerline{\includegraphics[width=0.5\textwidth]{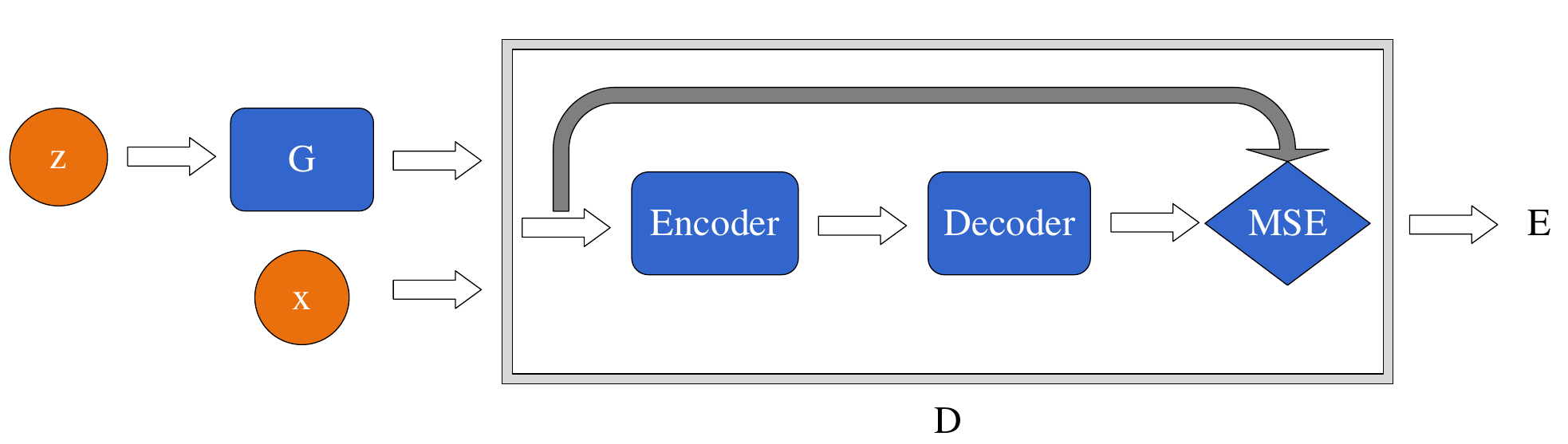}}
\caption{EBGAN autoencoder Model.}
\label{fig}
\end{figure}

EBGAN defines two different loss functions for the training of $G$ and $D$. Given a positive boundary $m$, a data sample $x$ and a generated sample $G(z)$, the discriminator loss $\mathcal{L}_D$ and generator loss $\mathcal{L}_G$ are as follows: 
\begin{equation}
\mathcal{L}_D(x,z)=D(x)+{[m-D(G(z))]}^+
\end{equation}
\begin{equation}
\mathcal{L}_G(z)=D(G(z))
\end{equation}
where $[\cdot]^+=max(0,\cdot)$.

In EBGAN autoencoder model, there is also the Pull-away Term operation, referred to as PT. This operation is imposed on the loss function of $G$ to avoid the model generating samples that are clustered in one or a few $p_{data}$ patterns. PT operates on a mini-batch orthogonal pairwise sample representation. Let $S\in\mathbb{R}^{s\times N}$  represent a batch of sample representations extracted from the encoder output layer. PT is defined as follows:
\begin{equation}
\begin{split}
f_{PT}(S)=\frac{1}{N(N-1)}\sum_i \sum_{j \neq i}(\frac{S_i^TS_j }{\left\|S_i\right\| \left\|S_j\right\|})^2
\end{split}
\end{equation}

\subsection{Spectral Normalization}
Spectral Normalization (SN) \cite{b14} is a technique to stabilize GAN training. The difficulty of GAN training lies in the control of discriminator. In high-dimensional spaces, the density evaluation of samples by discriminators is sometimes unreliable, which causes the generator to fail to learn a multimodal target distribution. SN stabilizes the training of the discriminator by restricting the weights to converge to a distribution. Zenati et al. \cite{b15} experimentally demonstrate that adding Lipschitz constraints to the discriminator in GANs can stabilize training. They also find that SN is also beneficial to regularize the encoder.

\subsection{NSL-KDD dataset}
NSL-KDD is a benchmark dataset for evaluating intrusion detection performance \cite{b27}. It consists of training set KDDTrain+ and test set KDDTest+. In NSL-KDD, each traffic sample consists of 41-dimensional features, of which there are 34 continuous features and 7 discrete features. By analyzing the meaning of 41-dimensional features in traffic records, these features are divided into 4 sets: Intrinsic, Content, Time-based and Host-based \cite{b28}. For a more vivid description, we take a traffic record in the dataset as an example, see Fig. 2. In addition, there are five categories of traffic records: Normal, DoS, Probe, U2R and R2L.
\begin{figure}[htbp]
\centerline{\includegraphics[width=0.5\textwidth]{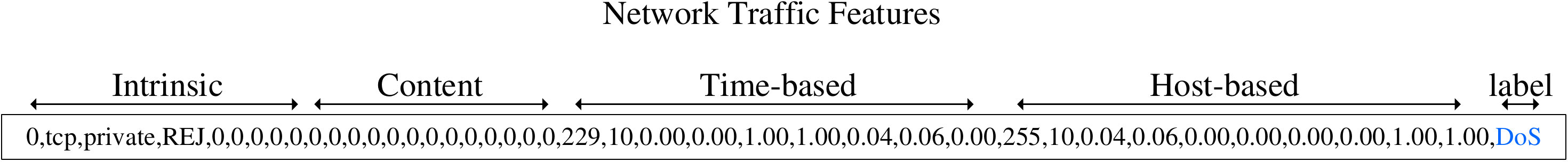}}
\caption{A network record in NSL-KDD.}
\label{fig}
\end{figure}

\section{Proposed Approach}

\subsection{Problem Definition}
Mathematically, we define our problem as follows: we train the model with the network traffic $D_{train}$ in KDDTrain+ and evaluate the model with the traffic record $D_{test}$ in KDDTest+. The traffic records in $D_{train}$ are divided into two categories according to their labels: normal traffic $D_{normal}$ and malicious traffic $D_{abnormal}$. For the samples in $D_{abnormal}$, $G$ in IDS-EBGAN adds different perturbations to the original traffic samples according to the class of attack to convert them to adversarial malicious traffic.

The IDS-EBGAN model diagram is shown in Fig. 3. Based on the datasets defined above, we train and test IDS-EBGAN. The training goal of the model is to learn the data distribution in $D_{normal}$. The network learns the features of normal traffic by minimizing the loss functions $\mathcal{L}_D$ and $\mathcal{L}_G$. Besides, training adversarial malicious network traffic can improve the detection rate of the discriminator for these abnormal traffic. Next, we define an anomaly function $A(\cdot)$, which yields a smaller anomaly score for normal network traffic and a larger score for malicious traffic. Therefore, for a given traffic sample $x$, the anomaly score $A(x)$ implies whether $x$ is normal or malicious.
\begin{figure*}[htbp]
\centerline{\includegraphics[width=0.8\textwidth]{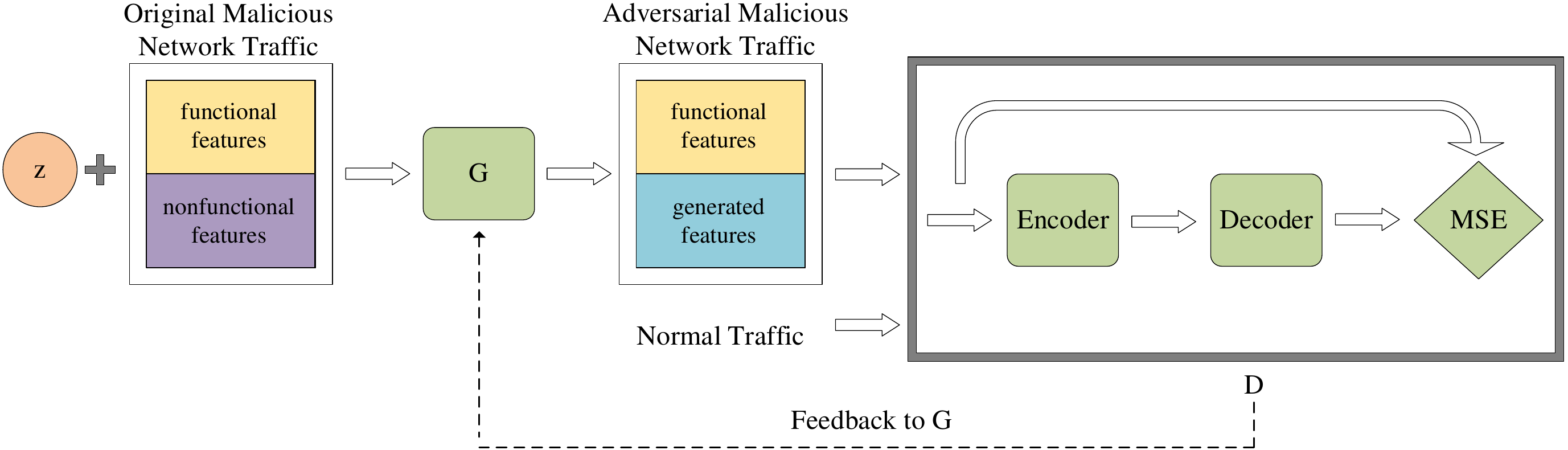}}
\caption{IDS-EBGAN.}
\label{fig}
\end{figure*}

\subsection{The Generation of Adversarial Malicious Traffic}\label{SCM}
The first priority in adversarial malicious traffic generation is to preserve the attack function of network traffic. There are four types of attacks in the NSL-KDD. Each class of attacks has their own functional and non-functional features. The functional characteristics of each type of attack represent the basic functions of this attack. Once the functional characteristics are changed, the network traffic no longer has the attack function. The following Table \Rmnum{1} describes the functional characteristics of each type of attack in the NSL-KDD dataset.
\begin{table}[htbp]
\caption{Relationship between attack categories and functional features}
\begin{center}
\begin{tabular}{|c|c|c|c|c|}
	\hline
	\multirow{2}{*}{Attack Type}&
	\multicolumn{4}{c|}{Functional Characteristics}\cr\cline{2-5}
		&Intrinsic&Content&Time-based&Host-based\cr
	\hline
	Probe& \checkmark & \quad & \checkmark &\checkmark \cr\hline
	DoS& \checkmark & \quad & \checkmark &\quad \cr\hline
	U2R& \checkmark & \checkmark & \quad &\quad \cr\hline
	R2L& \checkmark & \checkmark & \quad &\quad \cr
	\hline
\end{tabular}
\label{tab1}
\end{center}
\end{table}

$G$ in IDS-EBGAN ensures that its functional characteristics are not changed when generating a certain class of adversarial malicious traffic. In other words, we only change the non-functional characteristics of original malicious traffic. For example, when generating adversarial examples of the DoS category, the features belonging to Intrinsic and Time-based traffic are preserved, and we only change the features of the Content and Host-based traffic.

\subsection{Malicious Traffic Detection}
IDS-EBGAN is based on the idea of sample reconstruction. Normal network traffic can be reconstructed well, but malicious traffic is poorly reconstructed. Anomaly score is a very important indicator of the intrusion detection model, which can measure the similarity between the test sample and its reconstructed sample. We use the anomaly score to measure the abnormal degree of a sample $x$. Anomaly function in our model is defined as follows:

\begin{equation}
A(x)=MSE(x,Decoder(Encoder(x)))
\end{equation}
Where $MSE$ stands for $mean\_squared\_error$. The larger the value of $A(x)$, the more likely the traffic record is to be malicious. When the anomaly score $A(x)$ is greater than a certain threshold $\emptyset$, the traffic record $x$ is judged to be malicious. The process for computing $A(x)$ is shown in Algorithm 1. In the next part of the experiment, we will demonstrate in detail the effectiveness of the sample anomaly measurement method we selected.

\IncMargin{1em} 
\begin{algorithm}
    \SetAlgoNoLine
    \SetKwInOut{Input}{\textbf{Input}}
    \SetKwInOut{Output}{\textbf{Output}}     
    $D_{normal}, D_{abnormal} \gets get\_class\_data(D_{train})$ \\      
    IDS-EBGAN $\gets train\_model(D_{normal},D_{abnormal})$ \\
    \BlankLine  
    \Input{$x$ in $D_{test}$, $Encoder$, $Decoder$}
    \Output{$A(x)$, where $A$ is the anomaly function}

    1:procedure INFERENCE \\
    2:\quad $\tilde{z} \gets Encoder(x) \quad \rhd Encode \quad samples$ \\
    3:\quad $\tilde{x} \gets Decoder(\tilde{z})  \quad \rhd Reconstruct \quad samples$ \\
    4:\quad return $MSE(x,\tilde{x})$ \\
    5:end procedure\\

    \caption{IDS-EBGAN for Intrusion Detection\label{alg.1}}
\end{algorithm}
\DecMargin{1em}

There are usually two ways to set the threshold of anomaly score:
\begin{itemize}
\item Set based on the ratio of normal and malicious traffic samples. This method usually takes the top c\% of samples with the highest anomaly scores after calculating the anomaly scores of all samples, and marks them as malicious traffic.
\item Adjust the threshold setting according to the specific working conditions. The specific practice can use the maximum anomaly score obtained by training normal network traffic during model training as the threshold. During the test, if the abnormal score of the sample to be tested is higher than the threshold, it is determined as malicious traffic.
\end{itemize}

\section{Experiments}
We experiment on the NSL-KDD dataset. In the data preprocessing stage, One-Hot encoding is performed for the non-numeric features in the discrete features, and finally the 41-dimensional network traffic is converted into 121-dimensional. At the same time, the input vectors are normalized with Min-Max Normalization. In the experiment, we use a proportion-based approach to set the threshold of anomaly scores. After calculating the anomaly scores of all samples, the 44\% of network traffic records with the highest anomaly scores $A(x)$ are categorized as malicious traffic. The formula for Min-Max Normalization is as follows:

\begin{equation}
\begin{split}
x^{\prime}=\frac{x-x_{min}}{x_{max}-x_{min}}
\end{split}
\end{equation}
Where $x$ is the value before normalization. $x_{max}$ and $x_{min}$ respectively represent the maximum and minimum values of the feature in the dataset.

Since our problem is a binary classification problem, according to the settings, the generator can only generate a certain class of adversarial malicious traffic samples at a time. Therefore, in the experiments, the data used for model training actually consists of normal network traffic and malicious traffic examples of a certain attack in $D_{train}$. In addition, considering that the U2R and R2L attack categories have fewer traffic records and they have similar functional characteristics, the attacks of these two categories are considered as a type of attack. In the following experiments, we take the generation of DoS-type adversarial attack samples as an example to verify the performance of the IDS-EBGAN.

\subsection{Experimental Results}
We use Precision, Recall, and F1 score to evaluate our model. The results for all models in the table are averaged over twenty runs. In Table \Rmnum{2}, IDS-EBGAN achieves competitive results on NSL-KDD. Compared with AE, our model has obvious advantages. Python Outlier Detection (PyOD) \cite{b29} is the most popular Python anomaly detection tool library today. PyOD includes nearly 20 common anomaly detection algorithms. Except for EGBAD, ALAD, GANomaly and AE, the rest of the models are implemented using the algorithms in the PyOD.

\begin{table}[htbp]
\caption{Results On NSL-KDD}
\begin{center}
\begin{tabular}{|c|c|c|c|}
	\hline
	Model	 & Precision & Recall	 & F1 score \\
	\hline
	HBOS & 0.7502 & 0.7923 & 0.7807 \\
	IForest & 0.6802 & 0.8260 & 0.7462 \\
	KNN &0.6437 & 0.6620 & 0.6570 \\
	LOF & 0.6273 & 0.7405 & 0.6792 \\
	OCSVM	 & 0.6774 & 0.6818 & 0.6795 \\
	PCA & $0.7912$ & $0.7574$ & $0.8039$ \\
	AE & $0.9167$ & $0.9290$ & $0.9386$ \\
	EGBAD & $0.7260$ & $0.7582$ & $0.7372$ \\
	ALAD & $0.8309$ & $0.7998$ & $0.8001$ \\
	GANomaly & $0.8427$ & $0.8598$ & $0.8501$ \\
	\textbf{Ours with SN} & \textbf{0.9289} & \textbf{0.9317} & \textbf{0.9405} \\ 
	\hline
\end{tabular}
\label{tab1}
\end{center}
\end{table}

\subsection{Evaluation of Spectral Normalization}
Inspired by ALAD \cite{b15}, we add the SN to the encoder of the discriminator in IDS-EBGAN. Here, we also evaluate the performance impact of the addition of SN on intrusion detection. Similarly, for all the results in Table \Rmnum{3}, we have taken the average of twenty runs. From Table \Rmnum{3}, it is obvious that SN has a minor impact on the results of the experiment.
\begin{table}[htbp]
\caption{SN Performance}
\begin{center}
\setlength{\tabcolsep}{1.5mm}{
\begin{tabular}{|c|c|c|c|}
\hline
Model	 & Precision & Recall	 & F1 score \\
\hline
Ours with SN & 0.9289 & 0.9317 & 0.9405 \\
Ours without SN & $0.9280$ & $0.9321$ & $0.9399$ \\
\hline
\end{tabular}}
\label{tab3}
\end{center}
\end{table}

To further explore whether the addition of SN is beneficial to the training of discriminator, we record the loss value of discriminator during model training. In Fig. 4, it is evident that the loss value of the discriminator fluctuates less after SN is added.
\begin{figure}[htbp]
\centerline{\includegraphics{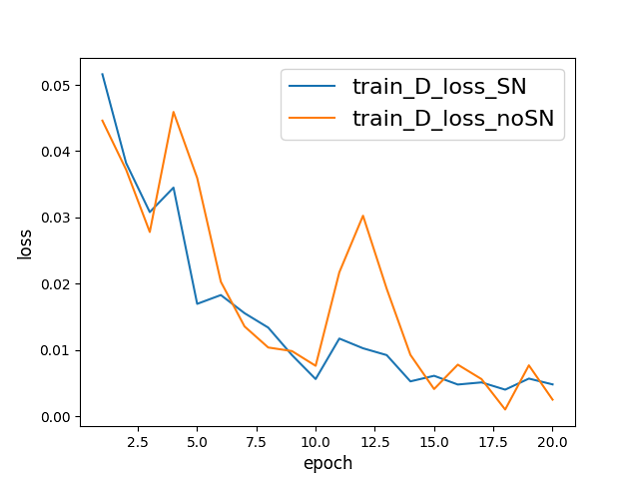}}
\caption{Effect of SN on discriminator training.}
\label{fig}
\end{figure}

\subsection{Evaluation of Anomaly Scores}
Here we evaluate our anomaly function and compare it with other criteria based on reconstruction. For example, we consider the $L_1$ reconstruction error and $MSE$ loss for comparison. Let $x$ be the input sample, $\tilde{x}=Decoder(Encoder(x))$ is the network traffic reconstructed by IDS-EBGAN.
\begin{center}
\begin{itemize}
      \item $L_1:A(x)=\left\|x-\tilde{x}\right\|_1$
 	\item $MSE:A(x)=MSE(x,\tilde{x})$
\end{itemize}
\end{center}

\begin{table}[htbp]
\caption{Different Anomaly Score}
\begin{center}
\begin{tabular}{|c|c|c|c|}
\hline
Model	 & Precision & Recall	 & F1 score \\
\hline
$L_1$ & $0.9013$ & $0.8901$ & $0.9067$ \\        
$MSE$ & $0.9289$ & $0.9317$ & $0.9405$ \\
\hline
\end{tabular} 
\label{tab4}
\end{center}
\end{table}

From the data in Table \Rmnum{4}, we can observe that $MSE$ performs better compared to the $L1$ reconstruction error. To more effectively evaluate the validity of the anomaly score we selected, during the model testing period, we calculate the anomaly scores of each network traffic in $D_{test}$ and obtain an anomaly score set $S$. Next, we use Min-Max normalization to convert the data in $S$ into [0,1] interval and obtain the set of anomaly scores $s^\prime$. From the anomaly score Histogram drawn from the data in $s^\prime$ (see Fig. 5), we can see that the score distribution of normal traffic and malicious traffic has a clear boundary, which shows the rationality of the anomaly score evaluation method we selected. 
\begin{figure}[htbp]
\centerline{\includegraphics{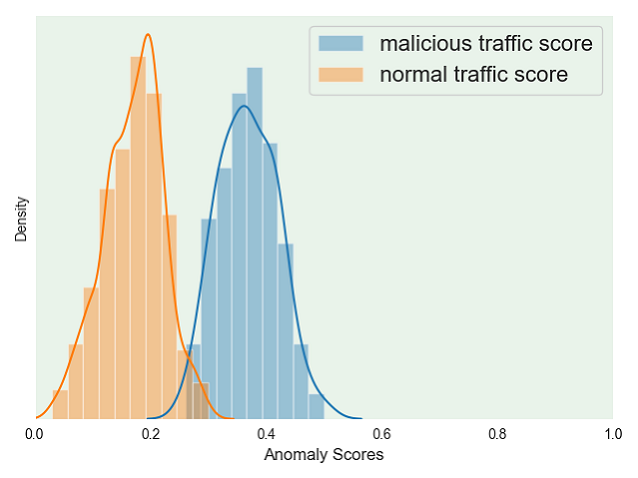}}
\caption{Histogram of anomaly scores for the test data.}
\label{fig}
\end{figure}

Furthermore, in the testing phase, we also cluster the reconstructed network traffic using KMeans \cite{b30} and visualize the results with t-SNE. Fig. 6 makes it clear that clustering results of reconstructed network traffic are clearly distinguished, which further confirms that our model can better reconstruct network traffic and distinguish them.

\begin{figure}[htbp]
\centerline{\includegraphics{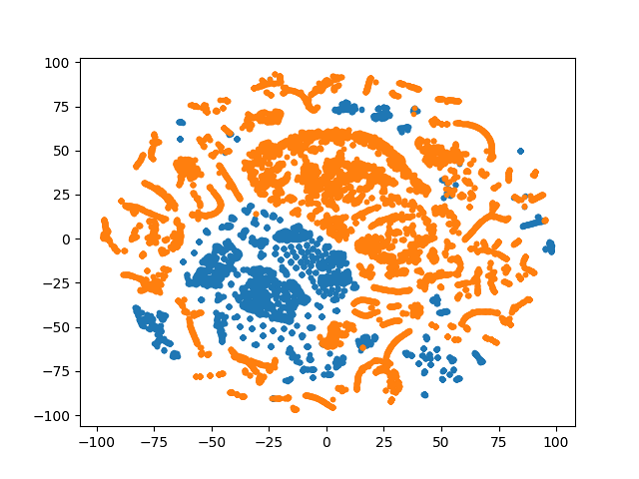}}
\caption{t-SNE visualization of the reconstruct samples.}
\label{fig}
\end{figure}

\subsection{The Role of Generator}
In IDS-EBGAN, the main role of $G$ is to generate adversarial malicious traffic. To study the impact of adversarial examples generated in the model on the intrusion detection task, we conduct relevant experiments. The input of the generator $G$ in IDS-EBGAN is the concatenation of the random noise vector and the original malicious traffic. The comparison experiment is to let the input of the $G$ be random noise vector, and the output is the fake network traffic instance. Similarly, we take the average of twenty results. As can be seen from Fig. 7, the generation of adversarial malicious traffic is of great significance to the model.
\begin{figure}[htbp]
\centerline{\includegraphics{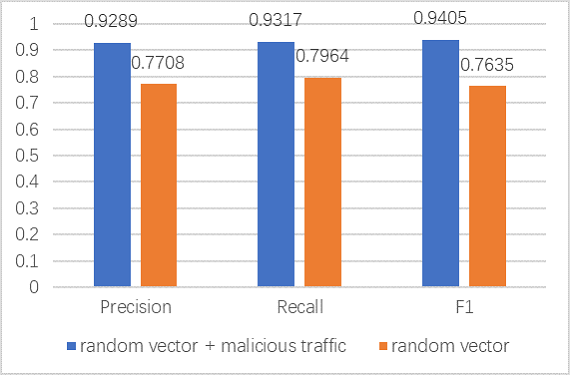}}
\caption{The performance of the generator.}
\label{fig}
\end{figure}

\subsection{Setting of Parameters}
The model is implemented using Pytorch1.8.1. In the experiments, the details of the architecture and hyperparameters of IDS-EBGAN are shown in Table \Rmnum{5}. Among them, we also explore the influence of the value of the positive boundary $m$ in the IDS-EBGAN loss function on the model. We conduct a wide range of experiments on the value of $m$ and find that the detection performance of the model is better when $m$=1.
\begin{table}[htbp]
\caption{Training Details}
\begin{center}
\begin{tabular}{l c c}
\hline
Operation	 & Units & Non-linearity \\
\hline
\multicolumn{3}{l}{$Generator(G)$} \\
\hline
\quad Linear & 512 & LeakyReLU \\
\quad Linear & 256  & LeakyReLU\\
\quad Linear & 23 & \quad \\
\hline
\multicolumn{3}{l}{$Discriminator-Encoder$} \\
\hline
\quad Linear & 512 & LeakyReLU \\
\quad Linear & 256 & LeakyReLU \\
\quad Linear & 100 & \quad \\
\hline
\multicolumn{3}{l}{$Discriminator-Decoder$} \\
\hline
\quad Linear & 128 & LeakyReLU \\
\quad Linear & 256 & LeakyReLU \\
\quad Linear & 121 & \quad\\
\hline
\multicolumn{3}{l}{$Hyperparameters$} \\
\hline
\quad Optimize & Adam & \quad \\
\quad Learning Rate & 0.0002 & \quad \\
\quad Batch Size & 64 & \quad \\
\quad Latent Dimensions & 100 & \quad \\
\quad Iterations & 20 & \quad \\
\quad Lambda\_pt & 0.1 & \quad \\
\quad $m$ & 1 & \quad \\
\hline
\end{tabular}
\label{tab5}
\end{center}
\end{table}

\section{Conclusion}
In this paper, an intrusion detection method based on EBGAN is proposed. In the training phase, the method utilizes the generation ability of the generator to produce adversarial malicious traffic while retaining the attack function of the traffic. After training, the discriminator can perfectly reconstruct normal network traffic, while the reconstruction loss of malicious traffic is relatively large. Through a series of experiments, we show that IDS-EBGAN is promising. Next, we will try to apply this method to other intrusion detection datasets, such as CIC-IDS-2018. In the future, we might try to use IDS-EBGAN in real working conditions to see its performance and make corresponding improvements.

\section*{Acknowledgment}
This work was supported by the Grand Joint Projects of Shanghai University grant 202124, Shanghai Engineering Research Center of Intelligent Computing System, Shanghai University (Grant number 19DZ2252600), The National Key Research and Development Program of China (No. 2018YFB0704400), Shanghai Sailing Program(20YF1414300), and GHfund B (Grand No.20210702).

\vspace{12pt}
\color{red}

\end{document}